\documentclass[conference]{IEEEtran}
\pdfoutput=1

\usepackage{xcolor}
\usepackage{mathtools}
\usepackage{amsmath}
\usepackage{amssymb}
\usepackage[vlined]{algorithm2e}
\usepackage{multirow}
\usepackage{subcaption}
\usepackage{pgfplots}
\usetikzlibrary{external}
\newcommand\doubleblind[1]{#1}
\newcommand\nda[1]{(missing due to NDA requirements)}

\IEEEoverridecommandlockouts
\IEEEpubid{\parbox{\columnwidth}{
\copyright2016 IEEE. Personal use of this material is permitted. Permission from IEEE must be obtained for all other uses, in any current or future media, including reprinting/republishing this material for advertising or promotional purposes, creating new collective works, for resale or redistribution to servers or lists, or reuse of any copyrighted component of this work in other works.\\%
Accepted for SC'16.
\hfill} \hspace{\columnsep}\makebox[\columnwidth]{ }}

\begin{document}

\title{The Vectorization of the Tersoff Multi-Body Potential: An Exercise in Performance Portability}
\author{
\doubleblind{
\IEEEauthorblockN{Markus H\"ohnerbach} \IEEEauthorblockA{RWTH Aachen University}
\and 
\IEEEauthorblockN{Ahmed E. Ismail} \IEEEauthorblockA{West Virginia University}
\and 
\IEEEauthorblockN{Paolo Bientinesi} \IEEEauthorblockA{RWTH Aachen University}}
}

\maketitle

\begin{abstract}
Molecular dynamics simulations, an indispensable research tool in computational chemistry and materials science, consume a significant portion of the supercomputing cycles around the world. We focus on multi-body potentials and aim at achieving performance portability. Compared with well-studied pair potentials, multibody potentials deliver increased simulation accuracy but are too complex for effective compiler optimization. Because of this, achieving cross-platform performance remains an open question. By abstracting from target architecture and computing precision, we develop a vectorization scheme applicable to both CPUs and accelerators. We present results for the Tersoff potential within the molecular dynamics code LAMMPS on several architectures, demonstrating efficiency gains not only for computational kernels, but also for large-scale simulations. On a cluster of Intel® Xeon Phi's, our optimized solver is between 3 and 5 times faster than the pure MPI reference. 
\end{abstract}

\IEEEpeerreviewmaketitle

%\tableofcontents
\section{Introduction}
Molecular dynamics (MD) simulations trace the trajectory of thousands, millions,
and even billions of particles over millions of timesteps, enabling materials
science research at the atomistic level. Such simulations are commonly run on
highly parallel architectures, and take up a sizeable portion of the computing
cycles provided by today's supercomputers. In this paper, we extend the LAMMPS
molecular dynamics simulator~\cite{Plimp}
with a new, optimized and portable implementation of the Tersoff multi-body potential~\cite{tersoff}.

+\IEEEpubidadjcol%

In many simulations, interactions among particles are assumed to occur
in a pair-wise fashion (particle-to-particle), as dictated by 
potentials such as the Coulomb or Lennard-Jones ones.
However, several applications in materials science require multi-body
potential formulations \cite{CompAspMany}.
With these, the force between two particles does not depend solely on their
distance, but also on the relative position of surrounding particles.
This added degree of freedom in the potential enables more accurate modeling but 
comes at the cost of additional complexity in its evaluation.
Because of this complexity, 
the optimization of multi-body potential is still largely unexplored.

By design, LAMMPS' main parallelization scheme is an MPI-based domain
  decomposition; this makes it possible to run on clusters and
supercomputers alike, and to tackle large scale systems.
Optionally, LAMMPS can also take advantage of shared-memory parallelism via OpenMP.
However, support for vectorization is limited to Intel architectures and
only to a few kernels.
Given the well-established and well-studied mechanisms for parallelism in MD 
codes~\cite{Plimp}, our efforts mostly focus on
vectorization, as a further step to fully exploit the computational power of
the available hardware.
On current architectures, SIMD processing contributes greatly to the system's peak performance;
SIMD units offer hardware manufacturers a way to multiply performance for compute-intense applications.
This principle is most pronounced in accelerators such as Intel's Xeon Phis and
Nvidia's Teslas.\footnote{
``Vectorization'' is not a term commonly used in the context of GPU-accelerated processing;
we will regardless use it also to describe the kind of data-parallelism that
exists within GPU's, otherwise usually referred to as a ``warp''. 
}

Several successful open-source MD codes---e.g.~Gromacs~\cite{gromphi}, NAMD, LAMMPS~\cite{user-intel}, and ls1
mardyn~\cite{ls1phi}---already
take advantage of vectorization in their most central kernels. However, these
vectorized kernels usually do not include multi-body potentials.
Implementation methods for the kernels vary between hand-written assembly,
intrinsics (compiler-provided functions that closely map to machine
instructions), and annotations that guide the compiler's optimization
\cite{pragma}. 
Furthermore, the majority of these kernels are not readily portable to different
architectures; on the contrary, for each target architecture, separate
optimizations are required.
Our objective is an approach sufficiently general to attain
high performance on a wide variety of architectures, and that requires
changes localized in few, general building blocks.
We identified such a solution for the Tersoff potential. 
When transitioning from
one architecture to another, the  numerical algorithm stays fixed,
and only the interface to the surrounding software components (memory
management, pre-processing) needs to be tailored.

We demonstrate our approach for performance portability on a range of
architectures, including ARM, Westmere to
Broadwell, Nvidia's Teslas (from the Kepler generation),
and two generations of Intel Xeon Phi (Knights Corner and Knights
Landing). As already mentioned, the numerical algorithm---built on top of
platform-specific building blocks---stays the same across all architectures; 
it is the building blocks that are implemented (once and for all) for each of the instruction sets.
Our evaluation ranges from single-threaded to
a cluster of nodes containing two Xeon Phi's.
Depending on architecture and benchmark, we report speedups ranging
from 2x to 8x.

\paragraph*{Related Work}

In addition to the aforementioned LAMMPS, several other MD
codes are available, including 
Gromacs~\cite{gromacs}, NAMD~\cite{namd}, DL\_POLY2~\cite{dl-poly}, and
ls1\_mardyn~\cite{ls1}.
Gromacs is well known for its single-threaded performance, 
provides support for the Xeon Phi~\cite{gromphi},
and already contains a highly portable scheme for the vectorization of 
pair potentials~\cite{gromacs-vect}.
All the other softwares also contain routines specific to certain platforms such as the Phi~\cite{ls1phi} or CUDA.

LAMMPS \cite{Plimp} is a simulator 
written in C++ and designed to favor extensibility.
It excels at scalability using MPI, and comes with a number of optional
packages to run on a different platforms with different parallel paradigms.
LAMMPS supports OpenMP shared-memory programming, GPUs~\cite{lammps-gpu}, KOKKOS~\cite{kokkos}, and vector-enabled Intel hardware~\cite{lammps-intel}.

The vectorization of pair-potential calculations was specifically addressed in an
MD application called
miniMD~\cite{pen} (proxy~\cite{miniapp} for the short-ranged
portion of LAMMPS). The target of that work are
various x86 instruction set extensions, including the Xeon Phi's IMCI, and
the optimization of cache access patterns.

There have been efforts to speed up pair potentials on GPUs~\cite{Pall20132641};
these techniques are similar to what one would use to achieve
vectorization, and 
the pattern of communication between the GPU and the host is similar to what
is needed to achieve high-performance with the %
first generation of
the Xeon Phi. %;
In general, GPUs have been used to great effect for speeding up MD
simulations~\cite{Anderson20085342,Rapaport2011926,Fan20131414}.

There exist implementations for GPUs that support
multi-body potentials such as EAM~\cite{lammps-intel},
Stillinger-Weber~\cite{gpub} and Tersoff~\cite{Hou20131364}. 
As opposed to our work, the Tersoff implementation for the GPU requires
explicit neighbor assignments and thus is only suitable for rigid problems; by
constrast, our approach is suitable for general scenarios.

In IMD, scalar optimizations similar to those we describe here
are implemented~\cite{imd0,imd1}; however, no sort of vectorization is included.
Our work is in part based on previous efforts to port
MD simulations, and LAMMPS in particular, to the Xeon Phi~\cite{lammps-intel};
specifically, we use the same data and computation management scheme.

\paragraph*{Organization of the paper}

In Sec.~\ref{sec:md-over}, we give a quick introduction to MD simulations
in general, while a discussion specific to the Tersoff potential and its
computational challenges comes in Sec.~\ref{sec:ters}.
We introduce our optimizations and vectorization schemes in 
Sec.~\ref{sec:opt}, and then describe the techniques to achieve portability in 
Sec.~\ref{sec:impl}.
In Sec.~\ref{sec:res}, we provide the results of our multi-platform evaluations,
and in Sec.~\ref{sec:conc} we draw conclusions.

\paragraph*{Open Source}

The associated code is available at \cite{gh}.

\section{Molecular Dynamics Background}\label{sec:md-over}

A typical MD simulation consists of a set of atoms (particles) and a sequence of timesteps.
At each timestep, the forces for each atom are calculated, 
and velocity and position are updated accordingly.
The forces are modeled by a potential $V(\mathbf{x})$ that depends solely on the positions $\mathbf{x}_i$ of each atom $i$.
$V$ represents the potential energy of the system;
the force on an atom $i$ is then the negative derivative of $V$
with respect to the atom's position $\mathbf{x}_i$ \cite{classical}:
\begin{equation}
\mathbf{f}_i = -\partial_{\mathbf{x}_i} V(\mathbf{x}).\label{eqn:pot-force}
\end{equation}

Most non-bonded potentials, such as Lennard-Jones and Coulomb, are pair
potentials: As such, they can be expressed as a double sum over the
atoms, where the additive term %($\phi$) 
depends only on the relative distance between the 
atoms:
\begin{equation}
V = \sum_i\sum_j \phi(r_{ij}).\label{eqn:pair-pot}
\end{equation}

In practice, Eq.~\ref{eqn:pair-pot} is computed by limiting the inner
summation ($j$) only to the set of atoms $\mathcal{N}_i$---known as the ``neighbor list''---that are within a
certain distance $r_C$ from atom $i$:
\begin{equation}
V = \sum_i\sum_{j \in\mathcal{N}_i} \phi(r_{ij}),\label{eqn:pot-neigh}
\end{equation}
\begin{equation}
\mathcal{N}_i = \{j : r_{ij} \leq r_C, j\neq i\}.\label{eqn:neigh}
\end{equation}
This simplification is based on the assumption 
that $\phi$ goes to zero as the distance $r$ increases. 
The assumption is valid for most pair potentials, 
even though some have to be augmented using long-ranged calculation schemes.
With this second formulation, the complexity for the computation of
$V$ decreases from quadratic (in the number of atoms) to linear, thus making large-scale simulations feasible.

Algorithm~\ref{algo:general-pair} illustrates how, based on a potential $V$ as given in Eq.~\ref{eqn:pot-neigh}, the potential energy $V$ and the
forces $F_i$ on each atom $i$ can be evaluated.
\begin{algorithm}
\caption{Calculation of potential and forces due to a pair potential.
}\label{algo:general-pair}
\For{$i$}
{
 \For{$j\in\mathcal{N}_i$}
 {
   $V \gets V + \phi(r_{ij})$\;
   $F_i \gets F_i - \partial_{x_i} \phi(r_{ij})$\;
   $F_j \gets F_j - \partial_{x_j} \phi(r_{ij})$\;
 }
}
\end{algorithm}

\section{The Tersoff potential}\label{sec:ters}

As opposed to pair potentials, multi-body potentials deviate from the form of Eq.~\ref{eqn:pair-pot}. 
In particular, $\phi$ is replaced by a term that depends on more than just the distance $r_{ij}$.
Instead, it might depend also on the distance of other atoms close to atom $i$
or $j$, and on the angle between $i$, $j$ and the surrounding atoms.
 
Omitting trivial definitions, the Tersoff potential~\cite{tersoff} 
is defined as follows:
\begin{align}
V & = \sum_i \sum_{j \in\mathcal{N}_i} \overbrace{f_C(r_{ij}) \left[ f_R(r_{ij}) + b_{ij}f_A(r_{ij}) \right]}^{V(i, j, \zeta_{ij})},\label{eqn:ters-1}\\
b_{ij} & = \vphantom{\sum_{r_j}}(1 + \beta^\eta\zeta_{ij}^\eta)^{-\frac{1}{2\eta}}, \eta\in\mathbb{R}\label{eqn:ters-2},\\
\zeta_{ij} & = \vphantom{\sum_{r_j}}\sum_{k\in\mathcal{N}_i\setminus\{j\}} \underbrace{f_C(r_{ik}) g(\theta_{ijk}) \exp(\lambda_3 (r_{ij} - r_{ik}))}_{\zeta(i, j, k)}.\label{eqn:ters-3}
\end{align}
Eq.~\ref{eqn:ters-1} indicates that two forces act between each pair of atoms $(i, j)$: an attractive force modeled by $f_A$, and a repulsive force modeled by $f_R$.
Both depend only on the distance $r_{ij}$ between atom $i$ and atom $j$.
The bond-order factor $b_{ij}$, defined by Eq.~\ref{eqn:ters-2}, however, is a scalar that depends
on all the other atoms $k$ in the neighbor list of atom $i$, by means of their distance $r_{ik}$, and angle $\theta_{ijk}$ via $\zeta_{ij}$ (from Eq.~\ref{eqn:ters-3}).
Since the contribution of the $(i, j)$ pair depends on other atoms $k$, Tersoff is a multi-body potential.
$f_C$ is a cutoff function, smoothly transitioning from 1 to 0;
$g$ describes the influence of the angle on the potential;
all other symbols in Eq.~\ref{eqn:ters-1}--\ref{eqn:ters-3} are parameters that were empirically determined by fitting to known properties of the modelled material.
Although these parameters mean that many lookups are necessary,
the functions within the potential ($f_R, f_A, f_C, g, \exp$) are
expensive to compute, thus making the Tersoff potential 
a good target for vectorization.

Eq.~\ref{eqn:ters-1}--\ref{eqn:ters-3} give rise to a triple summation;
this is mirrored by the triple loop structure of 
Algorithm~\ref{algo:ters-lammps}, which describes the implementation found in LAMMPS in terms of the functions $V(i, j, \zeta)$ and $\zeta(i, j, k)$, and calculates forces $F$ and potential energy $E$:
For all $(i, j)$ pairs of atoms, first $\zeta_{ij}$ is accumulated, and then the
forces are updated in two stages, first with the contribution of the $V(i, j, \zeta_{ij})$ term,
and finally with the contributions of the $\zeta(i, j, k)$ terms.

\begin{algorithm}
\caption{Calculation of potential energy and forces due to the Tersoff potential.
}\label{algo:ters-lammps}
\For{$i$}
{
 \For{$j\in\mathcal{N}_i$}
 {
  $\zeta_{ij} \gets 0$\;
  \For{$k\in\mathcal{N}_i\setminus\{j\}$}
  {
    $\phantom{F^{ij}_k}\mathllap{\zeta_{ij}} \gets \phantom{F^{ij}_k}\mathllap{\zeta_{ij}} +\, \zeta(i, j, k)$\;
  }
  $E \gets E + V(i, j, \zeta_{ij})$\;
  $F_i \gets F_i - \partial_{x_i} V(i, j, \zeta_{ij})$\;
  $F_j \gets F_j - \partial_{x_j} V(i, j, \zeta_{ij})$\;
  $\delta\zeta \gets \partial_{\zeta}V(i, j, \zeta_{ij})$\;
  \For{$k\in\mathcal{N}_i\setminus\{j\}$}
  {
    $\phantom{F_k}\mathllap{F_i} \gets \phantom{F_k}\mathllap{F_i} - \delta\zeta\cdot\,\partial_{x_i}\zeta(i, j, k)$\;
    $\phantom{F_k}\mathllap{F_j} \gets \phantom{F_k}\mathllap{F_j} - \delta\zeta\cdot\,\partial_{x_j}\zeta(i, j, k)$\;
    $F_k \gets F_k - \delta\zeta\cdot\,\partial_{x_k}\zeta(i, j, k)$\;
  }
 }
}
\end{algorithm}

For the following discussions, it is important to keep in mind the loop structure of Algorithm~\ref{algo:ters-lammps}:
It consists of an outer loop over all atoms (denoted by the capital letter $I$), 
an inner loop over  a neighbor list (denoted by the capital letter $J$),
and inside the latter, two more loops over the same neighbor list (denoted by the capital letter $K$).

As opposed to pair potentials, many-body potentials are used with extremely short neighbor lists $\mathcal{N}_i$.
In a representative simulation
run, $\mathcal{N}_i$ rarely contains more than four atoms.
Assuming that the size of $\mathcal{N}_i$ is $n$, the algorithm accesses 
the atoms in $\mathcal{N}_i$ a total of $2n^2$ times. 
In practice, constructing the neighbor list on every timestep would be too expensive.
Instead, the cutoff radius $r_C$ is extended by a so-called ``skin
distance''. Because atoms only move a certain distance per timestep, 
one can guarantee that no atom enters or exits the cutoff region for 
a certain number of timesteps by tracking all atoms also within the skin distance.
Consequently, the neighbor list also only needs to be rebuild after this many steps.
We denote the extended neighbor list by $\mathcal{S}_i$ instead of $\mathcal{N}_i$.
Given that the Tersoff potential incorporates a cutoff function $f_C$, the mathematical formulation is equivalent no matter if iterating through $\mathcal{N}_i$ or $\mathcal{S}_i$.
Nevertheless, as little computation as possible should be performed on skin atoms.
Efficiently excluding skin atoms is one of the major challenges for vectorization.

\section{Optimizations}\label{sec:opt}

This section discusses the various optimizations that we applied to the algorithm described in the previous section.
Some are inherited from the libraries that we integrate with (USER-INTEL and KOKKOS), such as optimized neighbor list build, time integration, and data access (e.g. alignment, packing, atomics).
These optimizations are generic in that they apply to any potential that uses that particular package.

We devised several other optimizations which are instead specific to the Tersoff
potential; they are detailed here.
\begin{enumerate}
\item Scalar optimizations. These improvements are useful whether one vectorizes or not.
We improve parameter lookup by reducing
  indirection, and eliminate redundant calculation by inlining function
  calls.
  This group of optimizations also includes the method described in
  Sec.~\ref{ssec:prec}, which aims to remove redundant calculations of the $\zeta$ term.

\item Vectorization. 
  We discuss details of our vectorization strategy in Sec.~\ref{ssec:vect}, where we present 
  different schemes, and describe their effectiveness for various vector lengths.

\item Optimizations that aid vectorization. As described in
  Sec.~\ref{ssec:avoid-mask} and~\ref{ssec:fil}, we aim at reduce the waste
  of computing resources on skin atoms. 
\end{enumerate}

\subsection{Pre-calculating Derivatives}\label{ssec:prec}

The first optimization we discuss consists in restructuring the algorithm so that $\zeta$
and its derivatives are computed only once, in the first loop, and the product
with $\delta\zeta$ is only performed in the second loop.
Since $\zeta$ and its derivatives---naturally---share terms, this modification has a measurable impact on performance.
Indeed, $\zeta$ can be calculated from intermediate results of the derivative evaluation at the cost of just one additional multiplication.

However, the computation of the derivatives in the first loop over $k$ requires additional storage:
While the derivatives with respect to the positions of atoms $i$ and $j$ can
be accumulated, the derivatives for $k$ have to be stored separately, as they belong to different $k$'s.
In our implementation, this list can contain up to a specified number $k_{max}$ of elements.
Should more than $k_{max}$ elements be necessary, the algorithm falls back to
the original scheme, thus maintaining complete generality.
Algorithm~\ref{algo:deriv} implements this idea.

\begin{algorithm}
\caption{Calculation of Tersoff potential and forces, taking into account skin atoms, pre-calculation derivatives, reverting to original approach after $k_{max}$ elements are stored.
}\label{algo:deriv}
\dots\;
  $\zeta_{ij} \gets 0$; $\partial_i\zeta\gets 0$; $\partial_j\zeta\gets 0$; $\forall k. \partial_k\zeta\gets 0$\;
  $\mathcal{K} \gets \{\}$\;
  \For{$k\in\mathcal{S}_i\setminus\{j\} \wedge |\mathcal{K}| < k_{max}$}
  {
    \If{$r_{ik} > r_C$}{ \textbf{continue}\;}
    $\phantom{\partial_k\zeta}\mathllap{\zeta_{ij}}      \gets \phantom{\partial_k\zeta}\mathllap{\zeta_{ij}} +\, \zeta(i, j, k)$\;
    $\phantom{\partial_k\zeta}\mathllap{\partial_i\zeta} \gets \phantom{\partial_k\zeta}\mathllap{\partial_i\zeta} + \partial_{x_i}\zeta(i, j, k)$\;
    $\phantom{\partial_k\zeta}\mathllap{\partial_j\zeta} \gets \phantom{\partial_k\zeta}\mathllap{\partial_j\zeta} + \partial_{x_j}\zeta(i, j, k)$\;
    $\phantom{\partial_k\zeta}\mathllap{\partial_k\zeta} \gets \phantom{\partial_k\zeta}\mathllap{\partial_k\zeta} + \partial_{x_k}\zeta(i, j, k)$\;
    $\mathcal{K} \gets \mathcal{K} \cup \{k\}$\;
  }

  \For{$k\in\mathcal{S}_i\setminus\{j\}\setminus\mathcal{K}$}
  {
    \dots\;
  }
  \dots\;
  $F_i \gets F_i - \delta\zeta\cdot\;\partial_i\zeta$\;
  $F_j \gets F_j - \delta\zeta\cdot\;\partial_j\zeta$\;
  \For{$k\in\mathcal{K}$}
  {
    $F_k \gets F_k - \delta\zeta\cdot\,\partial_k\zeta$\;
  }
  \For{$k\in\mathcal{S}_i\setminus\{j\}\setminus\mathcal{K}$}
  {
    \dots\;
  }

\dots\;
% }
%}
\end{algorithm}

\subsection{Vectorization Choices}\label{ssec:vect}

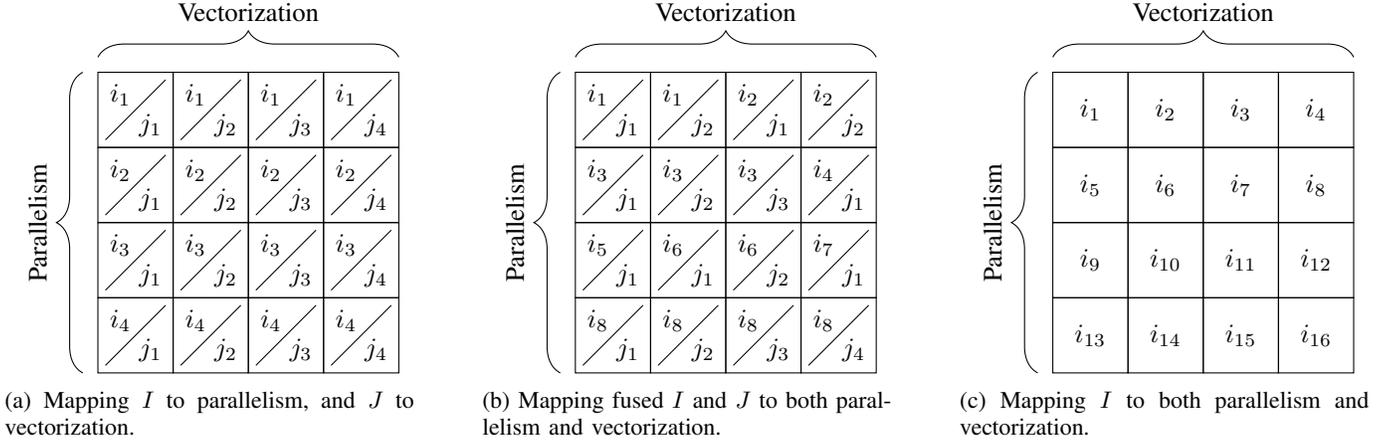
\begin{figure*}
\centering

\begin{subfigure}{.3\linewidth}
\centering

\begin{tikzpicture}

\newcommand{\myrect}[4]{
\draw (#1,#2) rectangle (#1+1,#2+1);
\draw (#1.1, #2.1) -- (#1.9, #2.9);
\node at (#1.3, #2.7) { #3 };
\node at (#1.7, #2.3) { #4 };
}

\myrect{1}{4}{$i_1$}{$j_1$}
\myrect{2}{4}{$i_1$}{$j_2$}
\myrect{3}{4}{$i_1$}{$j_3$}
\myrect{4}{4}{$i_1$}{$j_4$}

\myrect{1}{3}{$i_2$}{$j_1$}
\myrect{2}{3}{$i_2$}{$j_2$}
\myrect{3}{3}{$i_2$}{$j_3$}
\myrect{4}{3}{$i_2$}{$j_4$}

\myrect{1}{2}{$i_3$}{$j_1$}
\myrect{2}{2}{$i_3$}{$j_2$}
\myrect{3}{2}{$i_3$}{$j_3$}
\myrect{4}{2}{$i_3$}{$j_4$}

\myrect{1}{1}{$i_4$}{$j_1$}
\myrect{2}{1}{$i_4$}{$j_2$}
\myrect{3}{1}{$i_4$}{$j_3$}
\myrect{4}{1}{$i_4$}{$j_4$}

\draw [decorate,decoration={brace,amplitude=10pt}] (0.8,1.0) -- (0.8,5.0) 
 node [black,midway,xshift=-0.6cm,rotate=90] {Parallelism};
\draw [decorate,decoration={brace,amplitude=10pt}] (1.0,5.2) -- (5.0,5.2) 
 node [black,midway,yshift=0.6cm] {Vectorization};
\end{tikzpicture}

\caption{Mapping $I$ to parallelism, and $J$ to vectorization.}\label{sfig:vec-1}
\end{subfigure}%
\hfill
\begin{subfigure}{.3\linewidth}
\centering

\begin{tikzpicture}

\newcommand{\myrect}[4]{
\draw (#1,#2) rectangle (#1+1,#2+1);
\draw (#1.1, #2.1) -- (#1.9, #2.9);
\node at (#1.3, #2.7) { #3 };
\node at (#1.7, #2.3) { #4 };
}

\myrect{1}{4}{$i_1$}{$j_1$}
\myrect{2}{4}{$i_1$}{$j_2$}
\myrect{3}{4}{$i_2$}{$j_1$}
\myrect{4}{4}{$i_2$}{$j_2$}

\myrect{1}{3}{$i_3$}{$j_1$}
\myrect{2}{3}{$i_3$}{$j_2$}
\myrect{3}{3}{$i_3$}{$j_3$}
\myrect{4}{3}{$i_4$}{$j_1$}

\myrect{1}{2}{$i_5$}{$j_1$}
\myrect{2}{2}{$i_6$}{$j_1$}
\myrect{3}{2}{$i_6$}{$j_2$}
\myrect{4}{2}{$i_7$}{$j_1$}

\myrect{1}{1}{$i_8$}{$j_1$}
\myrect{2}{1}{$i_8$}{$j_2$}
\myrect{3}{1}{$i_8$}{$j_3$}
\myrect{4}{1}{$i_8$}{$j_4$}

\draw [decorate,decoration={brace,amplitude=10pt}] (0.8,1.0) -- (0.8,5.0) 
 node [black,midway,xshift=-0.6cm,rotate=90] {Parallelism};
\draw [decorate,decoration={brace,amplitude=10pt}] (1.0,5.2) -- (5.0,5.2) 
 node [black,midway,yshift=0.6cm] {Vectorization};
\end{tikzpicture}

\caption{Mapping fused $I$ and $J$ to both parallelism and vectorization.}\label{sfig:vec-2}
\end{subfigure}%
\hfill
\begin{subfigure}{.3\linewidth}
\centering

\begin{tikzpicture}

\newcommand{\myrect}[3]{
\draw (#1,#2) rectangle node { #3 } (#1+1,#2+1);
}

\myrect{1}{4}{$i_1$}
\myrect{2}{4}{$i_2$}
\myrect{3}{4}{$i_3$}
\myrect{4}{4}{$i_4$}

\myrect{1}{3}{$i_5$}
\myrect{2}{3}{$i_6$}
\myrect{3}{3}{$i_7$}
\myrect{4}{3}{$i_8$}

\myrect{1}{2}{$i_9$}
\myrect{2}{2}{$i_{10}$}
\myrect{3}{2}{$i_{11}$}
\myrect{4}{2}{$i_{12}$}

\myrect{1}{1}{$i_{13}$}
\myrect{2}{1}{$i_{14}$}
\myrect{3}{1}{$i_{15}$}
\myrect{4}{1}{$i_{16}$}

\draw [decorate,decoration={brace,amplitude=10pt}] (0.8,1.0) -- (0.8,5.0) 
 node [black,midway,xshift=-0.6cm,rotate=90] {Parallelism};
\draw [decorate,decoration={brace,amplitude=10pt}] (1.0,5.2) -- (5.0,5.2) 
 node [black,midway,yshift=0.6cm] {Vectorization};
\end{tikzpicture}
\caption{Mapping $I$ to both parallelism and vectorization.}\label{sfig:vec-3}
\end{subfigure}
\caption{Mapping of atoms ($I$) and elements of neighbor lists ($J$) to vector units and parallelism.}
\label{fig:map}
\end{figure*}

In Alg.~\ref{algo:deriv} (and Alg.~\ref{algo:ters-lammps})
the iteration space ultimately is
three-dimensional, corresponding to the three nested loops $I$, $J$ and $K$. 
This space needs to be mapped onto the available execution schemes, 
that is, data parallelism,  parallel execution, and sequential execution.
We propose three different mappings that are useful in different scenarios.
For all of them, it is convenient to map the $K$ dimension onto sequential execution,
because values calculated in the $K$ loop, the $\zeta$'s, are used in the surrounding $J$ loop, and data
computed in the $J$ loop, i.e. $\delta\zeta$, is then used within the second $K$ loop.

Therefore, the problem boils down to mapping the $I$ and $J$ dimensions onto a combination
of parallel execution, data parallelism, and if necessary, sequential execution.
In our reasoning,
we assume that the amount of available data-parallelism is unlimited.
In practice, the program sequentially executes chunks, and each chunk takes advantage of data parallelism.

As shown in Fig.~\ref{fig:map}, 
to perform the mapping sensibly
we propose three schemes:  
\begin{itemize}
\item[(\ref{sfig:vec-1})] $I$ is mapped to parallel execution, and $J$ to data parallelism.
\item[(\ref{sfig:vec-2})] $I$ is mapped to parallel execution, and $I$ and $J$ to data parallelism.
\item[(\ref{sfig:vec-3})] $I$ is mapped to parallel execution and data parallelism, and $J$ to sequential execution.
\end{itemize}

Scheme (\ref{sfig:vec-1}) is natural for vector architectures with short vectors, such as single precision SSE and double precision AVX.
In these, it makes sense to map the $j$ iterations directly to vector lanes, as there is a good match among them: 3-4 iterations to 4 vector lanes.
This scheme is most commonly used to vectorize pair potentials.
The advantage of this approach is that the atom $i$ is constant across all lanes.
While performing iterations in $k$ through the neighbor list of atom $i$, the same neighbor list is traversed across all lanes, leading to an efficient vectorization.
However, with long vectors and short neighbor lists, this approach is destined to fail on accelerators and CPUs with long vectors.

Scheme (\ref{sfig:vec-2}) is best suited for vector widths (8 or 16) that exceed
the iteration count of $j$,
as it handles the shortcomings of (\ref{sfig:vec-2}).  %\pdj{what is (2)?!?! careful with
With this approach, iterations of $i$ and $j$ are fused, and the fused loop is used for data parallelism.
Given that $i$ contains many iterations (as many as atoms in the system), this scheme
achieves an unlimited potential for data parallelism.
However, in contrast to (\ref{sfig:vec-1}), atom $i$ is not constant across all lanes;
consequently, the innermost loops iterates over the neighbor lists of
different $i$, leading to a more involved iteration scheme.
Even if this iteration is efficient, it can not attain the same performance of
an iteration scheme where all vector lanes iterate over the same neighbor list.
The vectorization of the $i$ loop invalidated a number of assumptions of the algorithm:
$i$ and $k$ are always identical across all lanes, while $j$'s, coming from the same neighbor list, are always distinct.
Without these assumptions, special care has to be taken when accumulating the forces to avoid conflicts.
For the program to be correct under all circumstances, the updates have to be serialized.
In the future, AVX-512 conflict detection support may change this.

Whether the disadvantages of Scheme (\ref{sfig:vec-2}) outweigh its advantages or not is primarily a question of amortization.
The answer depends on the used floating point data type, the vector length, and the features of the underlying instruction set.

Scheme (\ref{sfig:vec-3}) is the natural model for the GPU, where data
parallelism and parallel execution are blurred together. An iteration $i$ is
assigned to each thread, and the thread sequentially works through the $j$
iterations.

To implement these schemes, the algorithms are split into two components: a
``computational'' one, and a ``filter''.
The computational component carries out the numerical
calculations, including the innermost
loop and the updates to force and energy; the input to this component
are pairs of $i$ and $j$ for which the force and energy calculations are to be
carried out.
Given that the majority 
of the runtime is spent in computation, 
this is the part of the algorithm that has to be vectorized.
The filter component is instead responsible to feed work to the computational one;
its duty is to determine which pairs to pass.
To this end, the data is filtered to make sure that work is assigned to as many vector lanes
as possible before entering the vectorized part.
This means that the interactions outside of the cutoff region never even reach
the computational component.

\subsection{Avoiding Masking or Divergence}\label{ssec:avoid-mask}

\begin{figure}
\centering
    \begin{subfigure}[t]{0.5\linewidth}
        \centering
        \includegraphics[width=.5\linewidth]{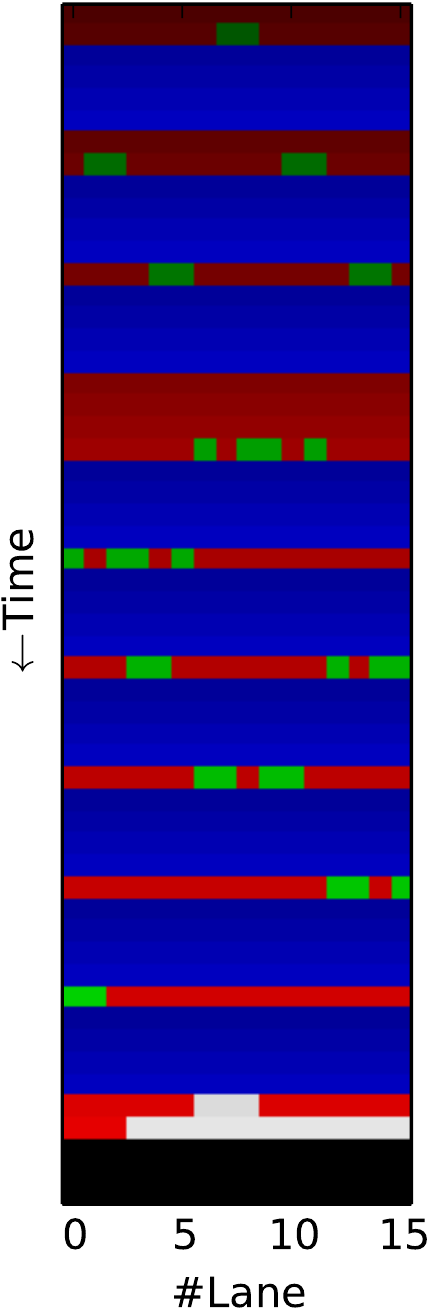}
    \end{subfigure}%
    ~ 
    \begin{subfigure}[t]{0.5\linewidth}
        \centering
        \includegraphics[width=.5\linewidth]{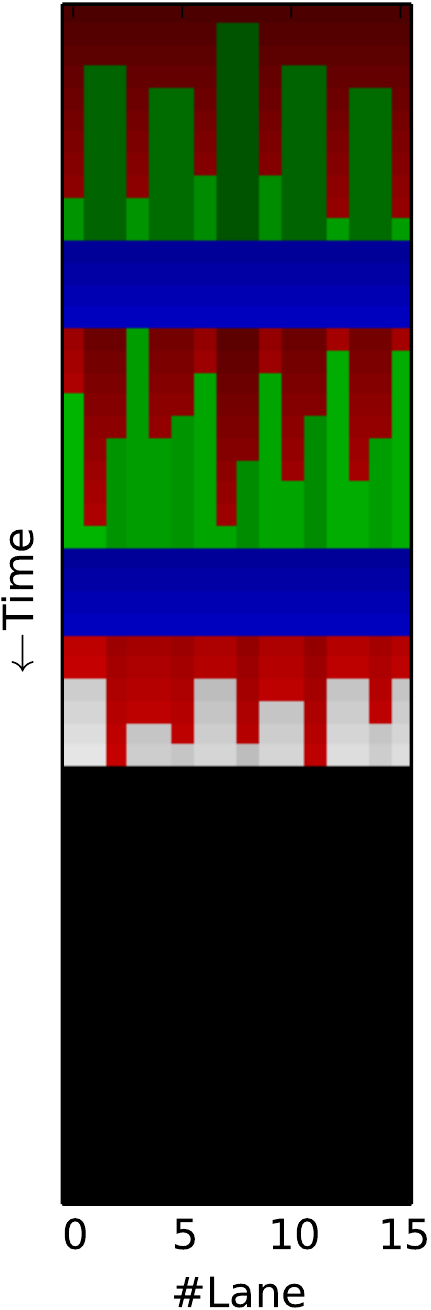}
    \end{subfigure}
\caption{Mask status during the iteration of the $K$ loop: Green is
  ready-to-compute, red is not-ready-to-compute, and blue is actual
  calculation. 
The left is an unoptimized variant, where calculation takes place as soon as at least one lane is ready-to-compute, whereas on the right, the calculation is delayed until all lanes are ready-to-compute.
}
\label{fig:avoid}
\end{figure}

Section~\ref{ssec:vect} just described how we avoid calculation for skin atoms in the $J$ loop.
The remaining issue is how we skip them in the $K$ loop.
The same argument, that resources must not be wasted on calculation that does not contribute to the final result, applies here, too.
As such, as many vector lanes as possible need to be active
before entering numerical kernels such as those computing
$\zeta(i, j)$ and  $V(i, j, \zeta)$. 
These computational kernels are almost entirely straight-line floating-point
intense code, with some lookups for potential parameters in
between. 

This optimization is most important for Scheme (\ref{sfig:vec-2}) and (\ref{sfig:vec-3}), as they traverse multiple neighbor lists in parallel.
As such, no guarantee can be made that interacting atoms have the same position in all neighbor lists.
This leads to sparse masks for the compute kernel:
For example, in a typical simulation that uses a vector length of sixteen,  
no more than four lanes will be active at a time.
On GPUs, this effect is even worse, where 95\% of the threads in a warp might be inactive.

Our optimization %\pdj{I so much dislike ``this'' by itself} 
extends Scheme (\ref{sfig:vec-2}) and (\ref{sfig:vec-3}) to fast forward through the $K$ loop, until as many lanes as possible can participate in the calculation of a numerical kernel.
The idea is that the neighbor list is not traversed at equal speed for all
the vector lanes; instead, we manipulate the iteration index independently in the various lanes.
Fig.~\ref{fig:avoid} visualizes the way this modification affects the behaviour of the algorithm.
In that figure, the shade of the particular color roughly corresponds to that lane's progress through the loop.
Notice that on the left, calculation (blue) takes place as soon as at least one lane requires that calculation (green).
Instead, on the right, a lane that is ready to compute (green), idles (does not change its shade), while the other lanes make further progress (going through shades of red) in search of the iteration where they become ready (green).
In our implementation, the calculation (blue) only takes place if all lanes are ready (green).
Effectively, we ``fast-forward'' in each lane, until all of them have are ready to compute.

\subsection{Filtering the Neighbor List}\label{ssec:fil}

To implement the idea from Sec.~\ref{ssec:avoid-mask},
a lot of masking is necessary, because the subset
  of lanes that have to progress when fast-forwarding changes every
  time.
On  platforms where masking has non-trivial overhead, performance can be further optimized.

Observe that in Fig.~\ref{fig:avoid}, lanes ``spin''  until computation is
available. 
We can reduce the amount of spinning by filtering the neighbor list in the scalar segment of the program.
To ensure correctness, the filtering is based on the maximum cutoff of all the types of atoms in the system.
This means that atoms that physically play a role 
can not be accidentally excluded from the calculation.
Filtering with any other cutoff might lead to incorrect results in systems with multiple kinds of atoms, if the cutoff prescribed between any two atom kinds differs.

Filtering the neighbor list is especially effective with AVX, where the double precision implementation uses the mapping (\ref{sfig:vec-1}), whereas the single precision variant uses (\ref{sfig:vec-2}).
Without this change, the overhead to spin  %\pdj{``to fast forward''} 
is too big to lead to speedups with respect to the double precision version.
With this change, most time is again spent in the numerical part of the algorithm.

\section{Implementation}\label{sec:impl}

So far we kept the description of the algorithms as generic as possible; in
this section, we cover the actual implementation of the schemes and optimizations from the previous section,
and their integration into LAMMPS.
From a software engineering standpoint, the main challenge was
to make the implementation maintainable, while achieving portable performance. 
To this end, 
OpenMP 4.5's SIMD extensions would be the most appealing solution, but right
now lack both compiler support
and a number of critical features that are required for our implementation.
We resorted to implementing these features ourself as modular and portable building blocks.
In the following,
we first introduce these building blocks, and 
then focus on a platform independent implementation. Finally, we characterize
the different execution modes supported by our code.

\subsection{Required Building Blocks}\label{ssec:openmp}

We identified four groups of 
building blocks necessary for a portable implementation.

(1) Vector-wide conditionals. These conditionals check if a condition is true
across all vector lanes; since either all or no lanes enter these conditionals,
excessive masking is prevented.

(2) Reduction operations. These are useful when all lanes accumulate to the same (uniform-across-lanes) memory location.
In these cases, the reduction can be performed in-register, and only the accumulated value is written to memory.
This behaviour can not be achieved with OpenMP's reduction clause,
since it only supports reductions in which the target memory location 
is known a-priori, while this is not the case for our code.

(3) Conflict write handling. 
This feature allows vector code to write to non distinct memory locations (see
the discussion in the previous bullet).
In the vectorization of MD codes, it is often guaranteed that all lanes
write to distinct memory locations (since the atoms in a
neighbor list are all distinct), which is the assumption that compilers typically make when performing user-specified vectorization.
Unfortunately, this guarantee does not hold for Scheme (\ref{sfig:vec-2}).
By serializing the accesses, 
the \texttt{ordered simd} clause of the OpenMP 4.5 standard provides a
solution to this issue; however, at time of writing, this directive is not yet supported by any major compiler.
It is also questionable whether this approach will be ``future proof'' or not, as 
a conflict detection mechanism such as that in the AVX-512 extensions might make serialization unnecessary.

(4) Adjacent gather optimizations. These provide improved performance on 
systems that do not support a native gather
instruction or where such an instruction has a high latency.
An adjacent gather is a sequence of gather operations that access adjacent memory locations.
Instead of using gather instructions or gather emulations here, it is possible to load continuously from memory into registers, and then permute the data in-register.
This operation can lead to significant performance improvements in our code, because adjacent gathers are necessary to load the parameters of our potential;
it is also important for backwards-compatibility reasons, because old
systems lack efficient native gather operations.

\subsection{Vector Abstraction}\label{ssec:veclib}

Since our objective is to integrate with the LAMMPS MD simulator,
support for different instruction sets and
for different floating point precisions is necessary.
It is crucial to support CPU instruction sets to balance the load between host and accelerator.
Additionally, such an abstraction
enables us to evaluate the influence of vector lengths and instruction set features on performance.
Considering all combinations of instruction sets, data types and vectorization
variants, it becomes clear that
it is infeasible to implement everything with intrinsics.

We created a single algorithm, and paired it with a vectorization back-end. 
As a consequence, instead of coding the Tersoff potential's algorithm
$n\cdot m$ times ($n$ architectures and $m$ precision modes), 
we only had to implement the building blocks for the vectorization back-end.
Some of these building blocks provide the features described in
Sec.~\ref{ssec:openmp}, while
others provide one-to-one mappings to intrinsics,
mostly  unary and binary operators.

The vectorization back-end uses C++ templates, which are
specialized for each targeted architecture.
We developed back-ends for single, double and mixed precision
using a variety of instruction set extensions:
Scalar, SSE4.2, AVX, AVX2, IMCI (the Xeon Phi Knights Corner instruction set), as well as experimental support for AVX-512, Cilk array notation and CUDA.

The library is designed to be easily extended to new architectures.
Even though the tuning might take some time, it is simplified by the fact that 
a number of building blocks, such as wide adjacent-gather operations, can be optimized in one go.
Contrary to most other vector libraries, which allow the programmer to pick a
vector length that may be emulated internally, our library only allows for
algorithms that are oblivious of the used vector length.

\subsection{Integration Points}\label{ssec:integrat}

We use vanilla LAMMPS' MPI-based domain decomposition scheme and
build upon optional packages that offer various optimizations and capabilities.
Specifically, all our x86 and ARM implementations use the USER-INTEL \cite{lammps-intel} package, which collects optimizations for Intel hardware, to manage offloading to the Xeon Phi, data-packing, alignment and simulation orchestration.
For the GPU implementation, the same role is fulfilled by the KOKKOS package
\cite{kokkos}.
Since KOKKOS abstracts the data layout of the variables used in a simulation
(e.g.~position, velocity, mass, force), the code needs to be changed wherever data is accessed from memory.
We also need to change the routine that feeds our algorithm, to conform with the model of parallelism that is used by KOKKOS.
As a consequence, comparisons between x86 and ARM, and between x86 and the GPU implementation cannot reasonably be drawn.
Furthermore, the KOKKOS package is still under development, while the USER-INTEL package is more mature;
as such, we believe that the GPU results are likely to have room for improvement.

\subsection{Accuracy}\label{ssec:acc}

\begin{figure}[t]
\centering
\includegraphics[width=\linewidth]{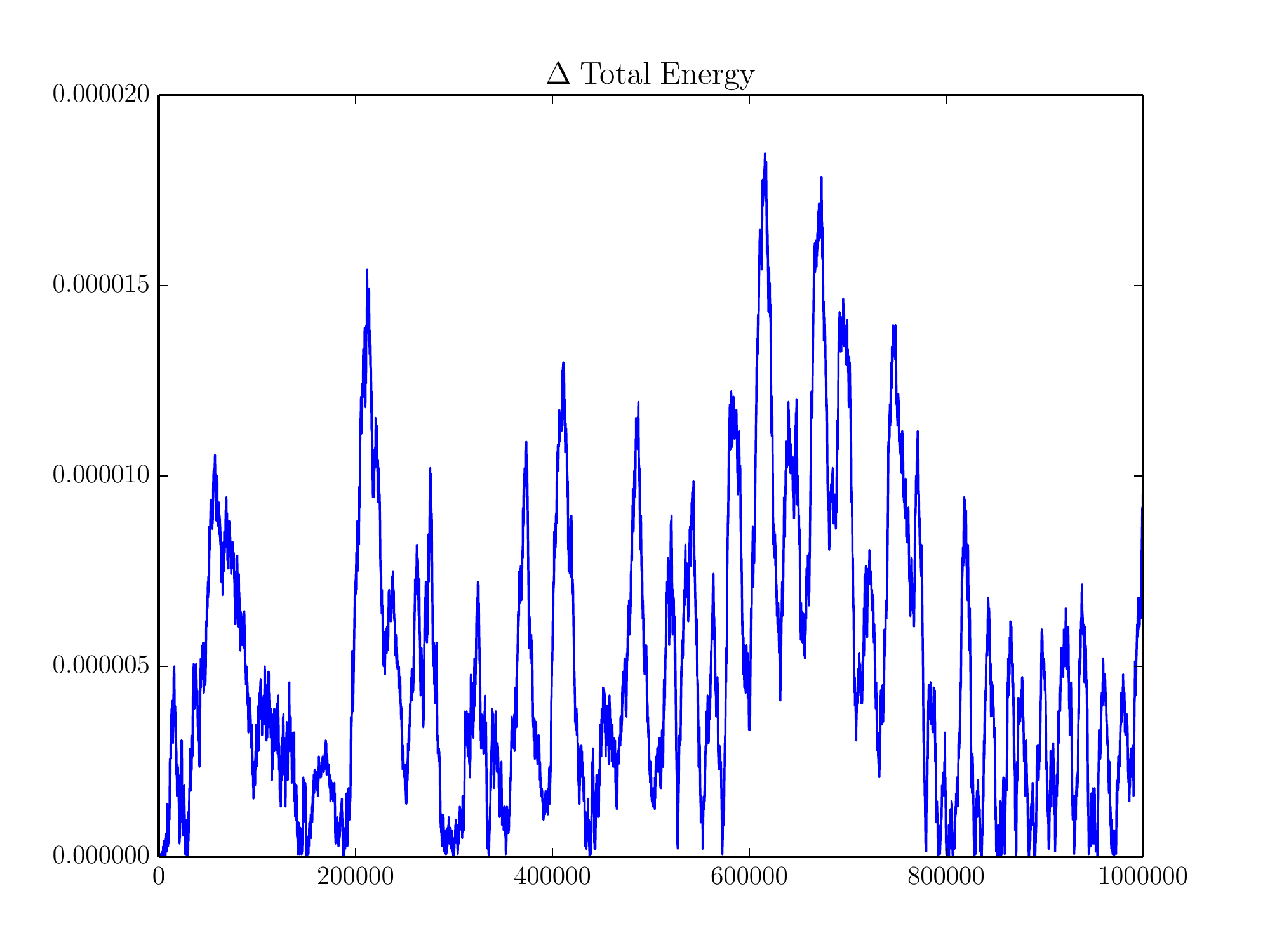}
\caption{Validation of the single precision solver: relative difference
  between the single precision and the double precision solvers for a system
  of 32.000 atoms, for one million of timesteps.}
\label{fig:acc}
\end{figure}

In addition to double precision, which is the default in LAMMPS,
we created versions that compute the Tersoff potential in single and mixed precision.
In order to validate these two implementations that use reduced precision, 
we measured the energy in a long-running simulation.
As Fig.~\ref{fig:acc} illustrates, for a system of 32000 atoms, the deviation is within 0.002\% of the reference.

In part, this effect can be explained by the short neighbor lists which
are characteristic of the Tersoff potential:
Since only few different atoms interact with any given atom, and only these accumulate their contributions to the force, there is little chance for round-off error to accumulate.

\subsection{Execution Modes for Measurement}\label{ssec:modes}

In the following section, we present performance results for several
hardware platforms, and four different codes:
Ref, Opt-D, Opt-S, Opt-M.

\paragraph*{Ref} 
The reference for our optimization and testing is the implementation shipped
with LAMMPS itself, which performs all the calculations in double precision.

\paragraph*{Opt-D}
The most accurate version of our code, which performs the calculations in double precision.
It includes both the optimizations due to scalar improvements, and those due to vectorization.

\paragraph*{Opt-S}
The least accurate version of our code, implemented entirely  in single precision.
As for Opt-D, it includes both scalar improvements, and takes advantage of vectorization.
The vector length
typically is twice that of Opt-D.
Referring to Sec.~\ref{ssec:acc}, the accuracy of the single precision solver is perfectly in line with the one offered by \emph{Opt-D} or \emph{Ref}. 

\paragraph*{Opt-M}
We also provide a mixed precision version of our code.
This version performs all the calculations in single precision, except for accumulations.
It is the default mode for code of the USER-INTEL package, as it offers a compromise between speed and accuracy.
From an software engineering perspective, the mixed precision version costs very little, as its can leverage the existing single and double precision codes;
indeed, our vector library performs this step (from single and double implementation to mixed implementation) automatically.

In addition to these four modes, the algorithms are run on
a single thread (\emph{1T}) or on an entire node (\emph{1N}).
The single-threaded run gives the most pure representation of the speedup
obtained by our optimizations; 
the results for an entire node (\emph{1N}) and a cluster instead give a realistic assessment of the expected speedup in real-world applications.
Such parallel runs use MPI, as provided by LAMMPS itself;
as a consequence, the parallelization scheme used in \emph{Ref} and \emph{Opt} is the same.

\section{Results}\label{sec:res}

In this section, we validate the effectiveness of our optimizations 
for the Tersoff potential
by presenting experimental results on a  variety of hardware platforms,
ranging from a low-power ARM to the second generation Xeon Phi.
As a test case, we use a standard LAMMPS benchmark for the simulation of
Silicon atoms; since the atoms are laid out in a regular lattice so that
each of them has exactly four nearest neighbors, this test case captures well
the scenario of small neighbor lists discussed in Sec.~\ref{sec:ters} and
Sec.~\ref{sec:opt}.

We start by presenting single-threaded and single-node results for the CPUs
(and the respective instruction sets) listed in Table~\ref{tbl:cpu-hw}; 
we continue with  measurements for two GPUs (Table~\ref{tbl:gpu-hw}),
and conclude with results for the Xeon Phi (Table~\ref{tbl:phi-hw}), in number of configurations.
We also present data for a cluster of nodes, to demonstrate the degree to
which the scalar and vector improvements lead to performance at scale. 

\begin{table}[h]
\caption{Hardware used for CPU benchmarks.}\label{tbl:cpu-hw}
\centering
\begin{tabular}{llll}
Name & Processor &  Cores & Vector ISA\\
\hline
ARM & ARM Cortex-A15 & 2 $\times$ 4\footnotemark{} & NEON\\
WM  & Intel Xeon X5675 & 2 $\times$ 6 & SSE4.2\\
SB  & Intel Xeon E5-2450 & 2 $\times$ 8 & AVX\\
HW  & Intel Xeon E5-2680v3 & 2 $\times$ 12 & AVX2\\
HW2 & Intel Xeon E5-2697v3 & 2 $\times$ 14 & AVX2\\
BW  & Intel Xeon E5-2697v4 & 2 $\times$ 18 & AVX2
\end{tabular}
\end{table}\footnotetext{little.BIG: Two different CPUs on a single chip, one powerful Cortex-A15, and a less powerful Cortex-A7. We will measure performance using only the Cortex-A15.}

\paragraph*{Timing Methodology} 
LAMMPS uses MPI timers to reports the time spent in various stages
throughout a simulation.
Our primary performance metric is ``simulated time over run time'', in
``nanoseconds per second'' (equivalent to ``iterations per seconds'').
All the timings %presented in this section 
exclude initialization and cleanup times, since
in any real-world simulation these stages are entirely negligible.
However, the timings do include all other stages, such as communication, data transfer,
neighbor list construction, and time integration.

\subsection{CPUs}

\begin{figure}
\begin{tikzpicture}
\begin{axis}[legend cell align=left,
    legend pos=north west,
    legend style={draw=none},
    ylabel={ns/day},
    ybar, bar width=7pt,
    enlarge x limits=0.25,
    ymin=0,
    symbolic x coords={ARM, WM, SB, HW},
    title={Single-Threaded Execution}
]
\addplot coordinates { (ARM, 0.135) (WM, 0.404) (SB, 0.455) (HW, 0.758) };
\addplot coordinates { (ARM, 0.323) (WM, 0.781) (SB, 1.562) (HW, 2.466) };
\addplot coordinates { (ARM, 0.862) (WM, 1.416) (SB, 2.003) (HW, 3.658) };
\addplot coordinates { (ARM, 0.000) (WM, 1.352) (SB, 2.003) (HW, 3.656) };
\legend{Ref-1T, Opt-D-1T, Opt-S-1T, Opt-M-1T}
\end{axis}
\end{tikzpicture}

\caption{Evaluation of performance portability across CPUs; single-threaded execution. 32000 atoms.}
\label{fig:cpu-single}
\end{figure}
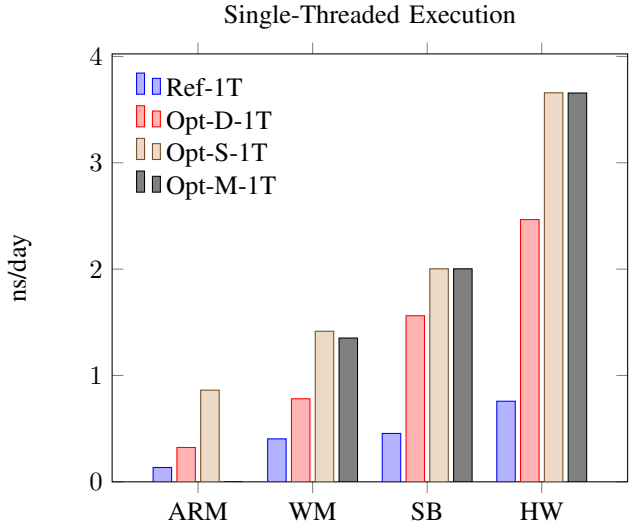

Fig.~\ref{fig:cpu-single} shows single-threaded performance 
for all the different execution modes described in Sec.~\ref{ssec:modes}.

On ARM, the speedups for Opt-D\footnote{Since NEON does not support
    vectorized double precision, Opt-D corresponds to optimized, but not
    vectorized, scalar version of the code.
    Due to lack of double precision vector instructions, we did not implement a mixed precision mode either.
}
 and Opt-S over Ref are 2.4 and 6.4, respectively.
These improvements can be partly attributed to alignment, to the use of lower accuracy math functions, and
  to our scalar optimizations; still, these influences do not explain the 2.7x
  improvement from Opt-S to Opt-D, which is mostly due to vectorization. 

On WM, we observe speedups of 1.9 between Ref and Opt-D,\footnote{
The SSE4.2 double precision results use the scalar back-end (since with a
vector length of two, vectorization does not yield speedups).
} and 3.5 between Ref and Opt-S. 
SSE4.2 supports vectorized integer instructions, so performance is limited by vector width and not features of the instruction set.

On SB (and HW), Opt-D is vectorized.\footnote{
For AVX/AVX2 in double precision (i.e.~SB/HW/BW), and SSE4.2 in single precision, we use scheme (\ref{sfig:vec-1}) from Section~\ref{ssec:vect}, whereas all longer vector lengths use the fused loop scheme (\ref{sfig:vec-2}).
}
This is readily apparent by the more than threefold speedup going from Ref to
Opt-D on SB.
Again, a portion of that speedup is due to scalar improvements, 
but it is clear that vectorization plays a significant role.
Since AVX lacks the integer instructions necessary to efficiently implement the (\ref{sfig:vec-2}) scheme from Section~\ref{ssec:vect}, the Opt-S/M measurements perform below expectations.

In contrast to SB, the HW systems utilizes the AVX2 instruction set and
attains a factor of 4.8 between Opt-S and Ref.
AVX2 adds integer and gather instructions, which our code takes advantage of.

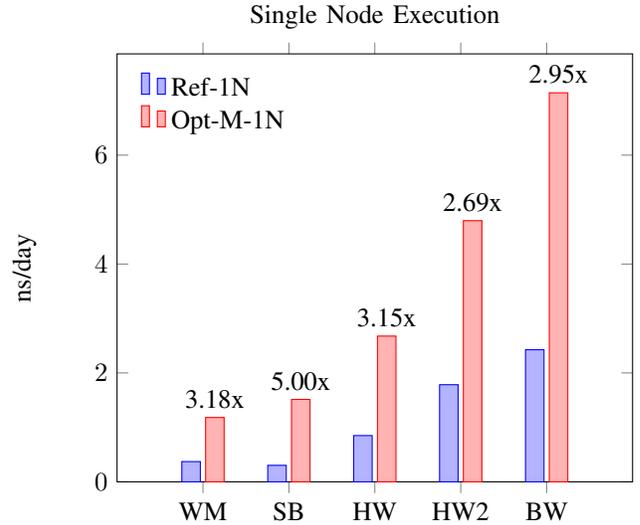
\begin{figure}
\begin{tikzpicture}
\begin{axis}[legend cell align=left,
    legend pos=north west,
    legend style={draw=none},
    ylabel={ns/day},
    ybar, bar width=7pt,
    enlarge x limits=0.25,
    ymin=0,
    symbolic x coords={WM, SB, HW, HW2, BW},
    nodes near coords,
    point meta=explicit symbolic,
    xtick={WM, SB, HW, HW2, BW},
    title={Single Node Execution}
]
\addplot coordinates { (WM, 0.372) (SB, 0.303) (HW, 0.850) (HW2, 1.784) (BW, 2.426) };
\addplot coordinates { (WM, 1.182) [\color{black}3.18x] (SB, 1.514) [\color{black}5.00x] (HW, 2.677) [\color{black}3.15x] (HW2, 4.797) [\color{black}2.69x] (BW, 7.145) [\color{black}2.95x] };
\legend{Ref-1N, Opt-M-1N}
\end{axis}
\end{tikzpicture}

\caption{Evaluation of performance portability across CPUs; one-node execution. 512000 atoms.}
\label{fig:cpu-all}
\end{figure}

In Fig.~\ref{fig:cpu-all}, we give results for the execution on entire nodes.
As these measurements use a bigger, more expensive simulation run, we only
  report on mixed precision measurements---mixed precision being the setting most likely to be used in practice.
The observed relative speedup between Ref and Opt-M ranges between 2.7 and 5.0;
these improvements are lower than what Fig.~\ref{fig:cpu-single} would suggest,
because the communication layer takes up between 5\% and 30\% of the execution
time.

\subsection{GPUs}

\begin{table}[t]
\centering
\caption{Hardware used for GPU benchmarks.}
\label{tbl:gpu-hw}
\begin{tabular}{l|lll|l}
Name & CPU & Cores & ISA & Accelerator\\
\hline
K20X  & Intel Xeon E5-2650 & 2 $\times$ 8 & AVX & Nvidia Tesla K20x\\
K40  & Intel Xeon E5-2650 & 2 $\times$ 8 & AVX & Nvidia Tesla K40\\
\end{tabular}
\end{table}

We now present results for two GPUs (listed in Table~\ref{tbl:gpu-hw}) from
the Kepler generation. As reference measurements, we use the LAMMPS GPU
package in double (Ref-GPU-D), single (Ref-GPU-S) and mixed (Ref-GPU-M)
precision, and the KOKKOS package in double precision (Ref-KK-D).  Our
optimized implementation Opt-KK-D is based on KOKKOS, and runs in double
precision.\footnote{ Following CUDA's paradigm, Opt-KK-D uses the scalar
  back-end, with one important distinction: The implementation of the
  vector-wide conditional operation uses a warp vote to determine if the
  condition is true for all threads in the warp.  This is semantically
  equivalent to the implementation for the vector instruction sets.  }

\begin{figure}
\hspace{-0.5em}%
\begin{tikzpicture}
\begin{axis}[legend cell align=left,
    legend pos=outer north east,
    legend style={draw=none},
    ylabel={ns/day},
    ybar,
    enlarge x limits=0.50,
    ymin=0,
    symbolic x coords={K20x, K40},
    xtick={K20x, K40},
    width=.8\linewidth,
    title={Offload to GPU}
]
\addplot coordinates { (K20x, 0.391) (K40, 0.441) };
\addplot coordinates { (K20x, 1.123) (K40, 1.204) };
\addplot coordinates { (K20x, 0.982) (K40, 1.071) };
\addplot coordinates { (K20x, 0.432) (K40, 0.485) };
\addplot coordinates { (K20x, 1.567) (K40, 1.767) };
\legend{Ref-GPU-D, Ref-GPU-S, Ref-GPU-M, Ref-KK-D, Opt-KK-D}
\end{axis}
\end{tikzpicture}
\caption{Evaluation of performance portability for Nvidia GPUs. 256000 atoms.}
\label{fig:gpu}
\end{figure}
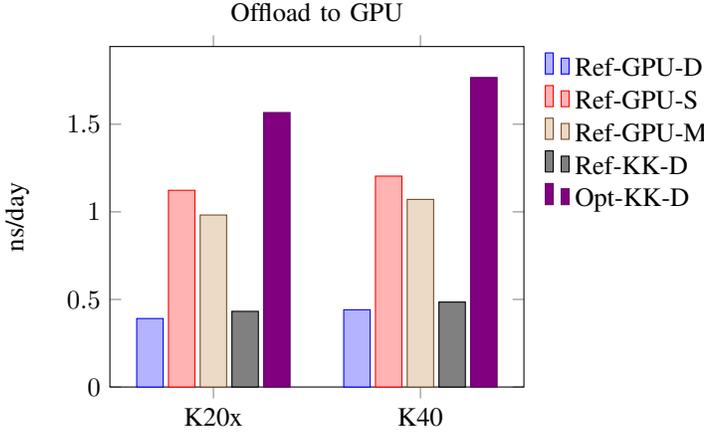

Fig.~\ref{fig:gpu} reports performance measurements when utilizing one entire GPU.
These measurements include very limited host involvement, to orchestrate the
offloading process; no force calculation, no communication, and no domain
decomposition takes place on the host.
This mode of running is probably most comparable to an ``entire host'' run from the previous section, as it utilizes the full GPU.
Even though the GPU programming paradigm is substantially different than the
CPU one, the speedup we achieve, roughly three, is similar.
While a single precision version of our optimized GPU code (Opt-KK-S) does not exist yet,
the reported measurements suggest that even higher performance (likely around 5ns/s) should be obtained.

In order to get a better idea of 
the benefits of the optimizations we performed, we look at 
Ref-KK-D and Opt-KK-D, and 
only compare the time for the routines affected by our optimizations 
(the others, such as communication and neighbor list builds, are identical for both versions). 
This isolated speedup is even higher, at approximately 5x.

\subsection{Intel Xeon Phi}

\begin{table*}
\centering
\caption{Hardware used in the evaluation of the Xeon Phi performance.}
\label{tbl:phi-hw}
\begin{tabular}{l|lll|lll}
Name & CPU & Cores & ISA & Accelerator & Cores & ISA\\
\hline
SB+KNC  & Intel Xeon E5-2450 & 2 $\times$ 8 & AVX & Intel Xeon Phi 5110P & 60 & IMCI\\
IV+2KNC  & Intel Xeon E5-2650v2 & 2 $\times$ 8 & AVX & Intel Xeon Phi 5110P & 2 $\times$ 60 & IMCI\\
HW+KNC  & Intel Xeon E5-2680v3 & 2 $\times$ 12 & AVX2 & Intel Xeon Phi 5110P & 60 & IMCI\\
KNL & -- & -- & -- & Intel Xeon Phi 7250 & 68 & AVX-512
\end{tabular}
\end{table*}

We conclude with a discussion of the portability of our optimizations
on two generations of Intel Xeon Phi accelerators---``Knights Corner'' (KNC),
and ``Knights Landing'' (KNL)---scaling from a single
accelerator to a cluster.

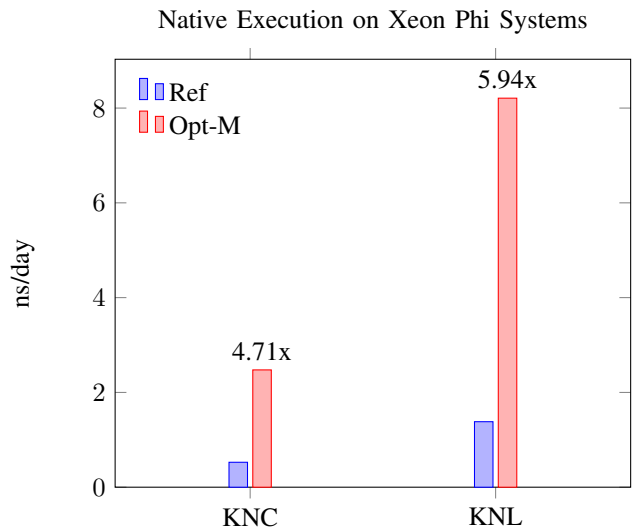
\begin{figure}
\begin{tikzpicture}
\begin{axis}[legend cell align=left,
    legend pos=north west,
    legend style={draw=none},
    ylabel={ns/day},
    ybar, bar width=7pt,
    enlarge x limits=0.55,
    ymin=0,
    xtick={KNC,KNL},
    nodes near coords,
    point meta=explicit symbolic,
    symbolic x coords={KNC,KNL},
    title={Native Execution on Xeon Phi Systems}
]
\addplot coordinates { (KNC, 0.526) (KNL, 1.382) };
\addplot coordinates { (KNC, 2.475) [\color{black}4.71x] (KNL, 8.209) [\color{black}5.94x] };
\legend{Ref,Opt-M}
\end{axis}
\end{tikzpicture}
\caption{Evaluation of performance portability for Xeon Phi's. 512000 atoms.}
\label{fig:speedup-phi}
\end{figure}

Fig.~\ref{fig:speedup-phi} measures the impact of our optimizations while using all cores of a Xeon Phi accelerator.
For a fair comparison, the benchmark is run on the device,
without any involvement of the host in the calculation.
On both platforms, the speedup of Opt-M with respect to Ref is
roughly 5x. Single-threaded measurements (not reported) 
indicate that the ``pure'' speedup in the kernel is even higher, at approximately 9x.

With a relative performance improvement of about 3x, 
the difference between KNC and KNL is in line with our expectations;
in fact, the theoretical peak performance also roughly tripled, along with the bandwidth, which roughly doubled.
We point out that no optimization specific to KNL was incorporated in our code;
the speedup was achieved by simply making sure that the vector abstraction 
complied with AVX-512.
Additional optimizations for KNL might take advantage of 
the AVX-512CD conflict detection instructions and 
different code for gather operations.

\begin{figure}
\begin{tikzpicture}
\begin{axis}[legend cell align=left,
    legend pos=north west,
    legend style={draw=none},
    ylabel={ns/day},
    ybar, bar width=7pt,
    enlarge x limits=0.25,
    ymin=0,
    symbolic x coords={SB+KNC, HW+KNC, IV+2KNC, KNL},
    title={Xeon Phi Performance},
    xticklabel style={font=\small}
]
\addplot coordinates { (SB+KNC, 2.029) (HW+KNC, 4.358) (IV+2KNC, 6.184) (KNL, 8.209) };
\legend{Opt-M}
\end{axis}
\end{tikzpicture}
\caption{Performance gains with Xeon Phis. 512000 atoms.}
\label{fig:nodes-phi}
\end{figure}
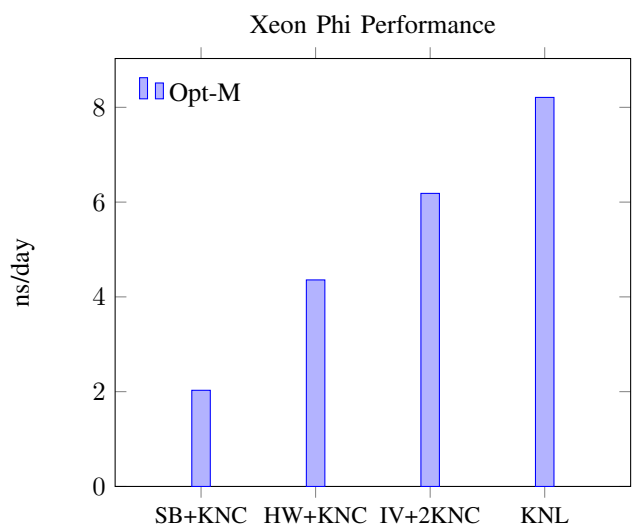

To lead up to the scaling results across multiple Xeon Phi augmented nodes, Fig.~\ref{fig:nodes-phi} measures the performance of individual such nodes as listed in Table~\ref{tbl:phi-hw}.
Like in a real simulation, the workload is shared among CPU and accelerator.
Given that our KNL system is self-hosted, we include it in this list.
The measurements for CPU+KNC, include both the overheads incurred in a single
node execution (such as MPI and threading), and the overhead due to offloading.
In view of the performance of the CPU-only systems relative to KNC, these performance numbers are then plausible.
A single KNC delivers higher simulation speed than the CPU-only SB node;
however, a CPU-only HW node is more powerful than the KNC, and thus also noticeably contributes to the combined performance.
Adding a second accelerator also seems to improve performance, as seen in the IV+2KNC measurement.
The KNL system delivers higher performance than the combination of two first-generation Xeon Phis and two Ivy Bridge CPUs.

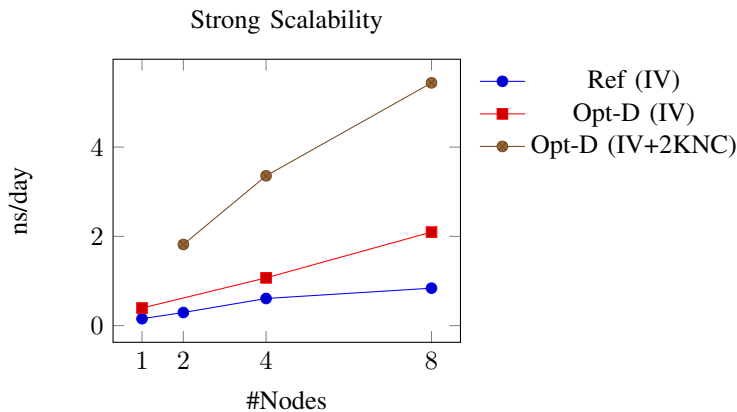
\begin{figure}
\centering
\begin{tikzpicture}
\begin{axis}[legend cell align=center,
    legend pos=outer north east,
    legend columns=1,
    legend style={draw=none},
    xtick={1,2,4,8},
    xlabel={\#Nodes},
    ylabel={ns/day},
    width=.7\linewidth,
    title={Strong Scalability}
]
\addplot coordinates { (1, 0.155) (2, 0.294) (4, 0.608) (8, 0.839) };
\addplot coordinates { (1, 0.394) (4, 1.070) (8, 2.096) };
\addplot coordinates { (2, 1.818) (4, 3.356) (8, 5.439) };
\legend{Ref (IV), Opt-D (IV), Opt-D (IV+2KNC)}
\end{axis}
\end{tikzpicture}
\caption{Optimization results on SuperMIC, a Xeon-Phi-augmented cluster. 2 million atoms.}
\label{fig:supermic}
\end{figure}

The question with any kind of serial improvement is
``will it translate to similar speedups in a (highly) parallel environment?''.
In theory, sequential improvements multiply with the performance achieved from parallelism;
in practice, a good chunk of those improvements are eaten away by a collection of overheads, and a realistic assessment can only be made from measurement.
Figure~\ref{fig:supermic} depicts results for up to eight nodes in a cluster of IV+2KNC nodes.
Here, overheads are not only due to the parallelism within a node, but also to
communication among the nodes.
The vector optimizations port to large scale computations seamlessly:
without accelerator, the performance improvement for 196 MPI ranks is 2.5x;
when two accelerators are added per node, the performance improvement becomes 6.5x.

\section{Conclusions}\label{sec:conc}

We discussed the problem of calculating the Tersoff potential efficiently and
in a portable manner; we described a number of optimization schemes, and validated their 
effectiveness by means of realistic use cases.
We showed that vectorization can achieve considerable speedups also in
scenarios---such as a multi-body potential with a short neighbor list---which 
do not immediately lend themselves to the SIMD paradigm.
To achieve portability, it proved useful to isolate target-specific code into a library of abstract operations; this separation of concerns leads to a clean division between the
algorithm implementation and the hardware support for vectorization.
It also makes it possible to map the vector paradigm 
to GPUs, while attaining considerable speedup.
The ideas behind our optimizations were described, and their effectiveness was
validated by means of realistic use cases.
Indeed, we observe speedups between 2x and 3x on most CPUs, 
and between 3x and 5x on accelerators; performance scales also to clusters and
clusters of accelerators.
Finally, we believe that the main success of this work lies in the achieved
degree of cross-platform code reuse, and in the portability of the
proposed optimizations; 
combined, these two features lead to a success story with respect to performance portability.

\section{Acknowledgments}

\doubleblind{
The authors gratefully acknowledge financial support from the Deutsche Forschungsgemeinschaft (German Research Association) through grant GSC 111, and from Intel via the Intel Parallel Computing Center initiative.
We thank the RWTH computing center and the Leibniz Rechenzentrum M\"unchen for computing resources to conduct this research.
We would like to thank Marcus Schmidt for providing one of the benchmarks used in this work, and M. W. Brown for conducting the benchmarks on the 2nd generation Xeon Phi hardware.
}

\end{document}